\documentclass[a4paper,11pt]{article}
\usepackage{pos}

\usepackage{caption,subcaption}

\pdfoutput=1

\let\OLDthebibliography\thebibliography
\renewcommand\thebibliography[1]{
	\OLDthebibliography{#1}
	\setlength{\parskip}{0pt}
	\setlength{\itemsep}{0pt plus 0.3ex}
}

\title{MeVCube: a CubeSat for MeV astronomy}
 \ShortTitle{MeVCube: a CubeSat for MeV astronomy}

\author*{Giulio Lucchetta}

\affiliation{DESY Zeuthen,\\
  Platanenallee 6, 15738, Zeuthen, Germany}

\emailAdd{giulio.lucchetta@desy.de}

\abstract{Despite the impressive progresses achieved both by X-ray and gamma-ray observatories in the last decades, the energy range between $\sim 200\,\mathrm{keV}$ and $\sim 50\,\mathrm{MeV}$ remains poorly explored. \emph{COMPTEL}, on-board \emph{CGRO} (1991-2000), was the last telescope to accomplish a complete survey of the MeV-sky with a relatively modest sensitivity. Missions like \emph{AMEGO} have been proposed for the future, in order to fill this gap in observation; however, the time-scale for development and launch is about 10 years. On a shorter time-scale, a different approach may be profitable: MeV observations can be performed by a Compton telescope flying on a CubeSat.\\
\emph{MeVCube} is a 6U CubeSat concept currently under investigation at \emph{DESY}, that could cover the energy range between hundreds of keV up to few MeVs with a sensitivity comparable to that of missions like \emph{COMPTEL} and \emph{INTEGRAL}. The Compton camera is based on pixelated Cadmium-Zinc-Telluride (CdZnTe) semiconductor detectors, coupled with low-power read-out electronics (ASIC, VATA450.3), ensuring a high detection efficiency and excellent energy resolution. In this work I will show measurements of the performance of a custom design CdZnTe detector and extrapolations of the expected telescope performance based on these measurements as well as simulations.}

\FullConference{37$^{\rm{th}}$ International Cosmic Ray Conference (ICRC 2021)\\
		July 12th -- 23rd, 2021\\
		Online -- Berlin, Germany}
		
\begin{document}
\maketitle

\section{Introduction}
\label{sec:intro}
\begin{figure}[ht]
\centering
\includegraphics[scale=0.40]{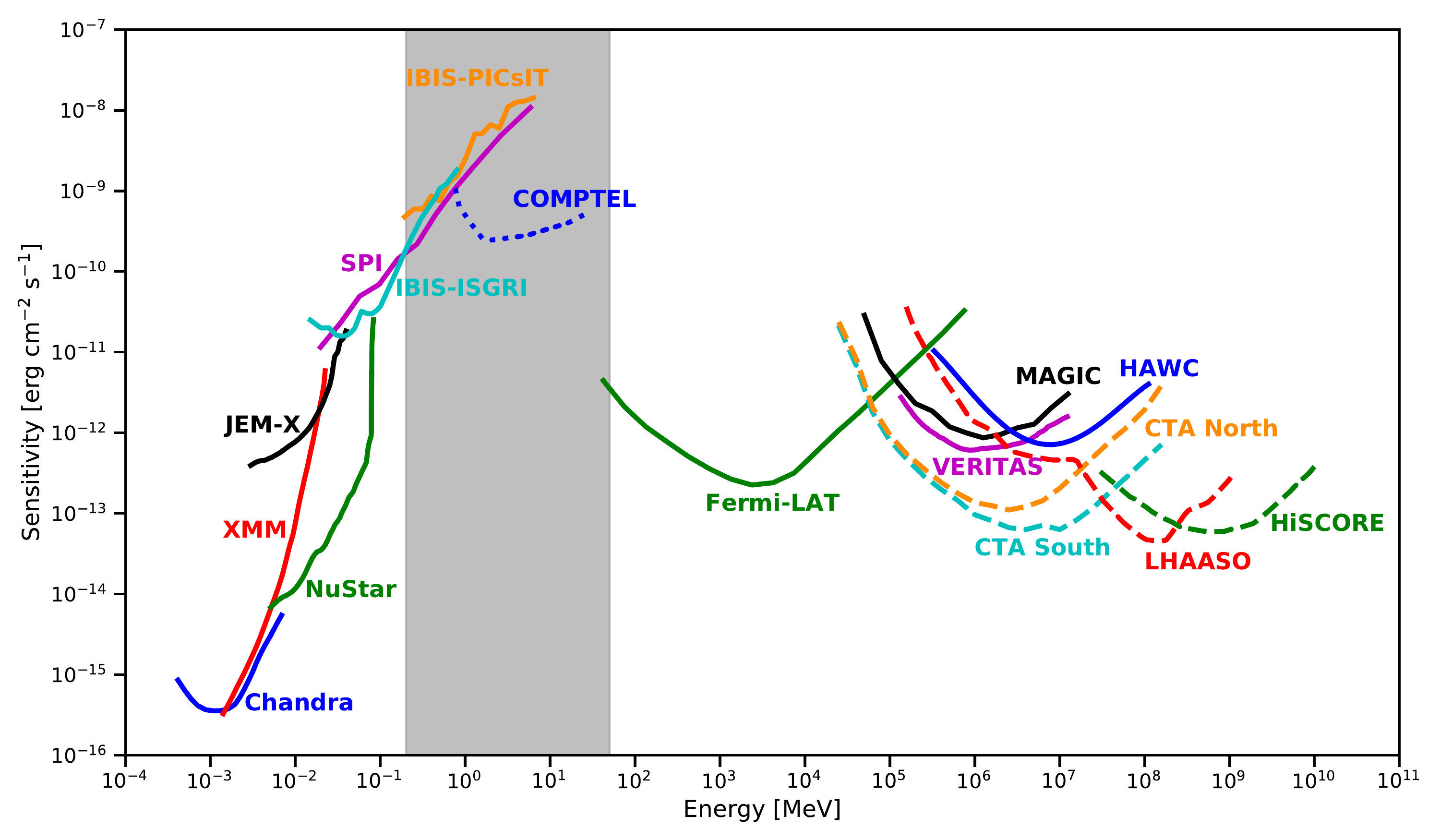}
\caption[MeVGap]{Differential sensitivity of different X-ray and $\gamma$-ray instruments. Past missions are shown in dotted lines, currently operating instruments in solid lines and future planned missions in dashed lines. Figure taken from Ref.~\citenum{Lucchetta_SPIE}.}
\label{fig:MeVGap}
\end{figure}

\noindent The measurement of X-ray and gamma-ray emission from astrophysical sources has revolutionized our understanding of the most energetic, non-thermal processes in the Universe. Despite the impressive progresses achieved both by X-ray and gamma-ray observatories in the last decades (see, for example, Ref.~\citenum{ReviewX-ray} \footnote{\url{https://indico.ict.inaf.it/event/720/}} and Ref.~\citenum{ReviewGamma-ray} for a review), the energy range between $\sim 200\,\mathrm{keV}$ and $\sim 50\,\mathrm{MeV}$ remains poorly explored (see Fig.~\ref{fig:MeVGap}). \emph{COMPTEL}, on-board \emph{CGRO} (1991-2000) \cite{Schoenfelder_COMPTEL}, was the last satellite to accomplish a complete survey of the MeV sky.\\
The scientific case for a deeper exploration of this energy range is strong and includes, for example, the search of electromagnetic counterparts of gravitational wave or neutrino events, the study of emission mechanisms in active galactic nuclei, searches of electron-positron annihilation features and nucleosynthesis of heavy elements. A comprehensive overview of the science case can be found in Ref.~\citenum{DeAngelis_eAstrogam}.\\
Missions have been proposed for the future, in order to fill this gap in observation, most notably AMEGO \cite{Kierans_AMEGO}. However, the time-scale for development and launch of a probe class mission is around 10 years. On a shorter time-scale \emph{COSI} \cite{Tomsick_COSI} have been proposed as a small explorer mission. Looking at this scenario, a different approach may be profitable: MeV observations can be performed by a Compton camera, flying on a nano-satellite telescope, based on the CubeSat standard. The small cost and relatively short development time are clear advantages. Moreover such instrument could also be used as a pathfinder mission, to test technologies and algorithms to be employed in future large-scale missions. In this work the performance of a CubeSat MeV telescope, called \emph{MeVCube}, is evaluated: Sec.~\ref{sec:cubesat} introduces the CubeSat design and expected telescope performance based on simulations, while experimental measurements on a custom design detector and read-out electronics are shown in Sec.~\ref{sec:results}.
 
\section{MeVCube design and performance}
\label{sec:cubesat}
\noindent The term CubeSat indicates a class of nano-satellites with a standardized form factor: the standard unit ($1\,\mathrm{U}$) has a volume of  $10 \times 10 \times 11.35\;\mathrm{cm}^3$ and a maximum weight of $1.33\;\mathrm{kg}$. More than one unit can be combined together and the current CubeSat Design Specification defines the envelopes for $1\,\mathrm{U}$, $1.5\,\mathrm{U}$, $2\,\mathrm{U}$, $3\,\mathrm{U}$ and $6\,\mathrm{U}$ form factors \cite{CubeSatSpecification_2015}, with possible extensions to $12\,\mathrm{U}$ and $16\,\mathrm{U}$.\\

\begin{figure}[ht]
\begin{center}
\begin{tabular}{c} 
\includegraphics[width=0.95\columnwidth]{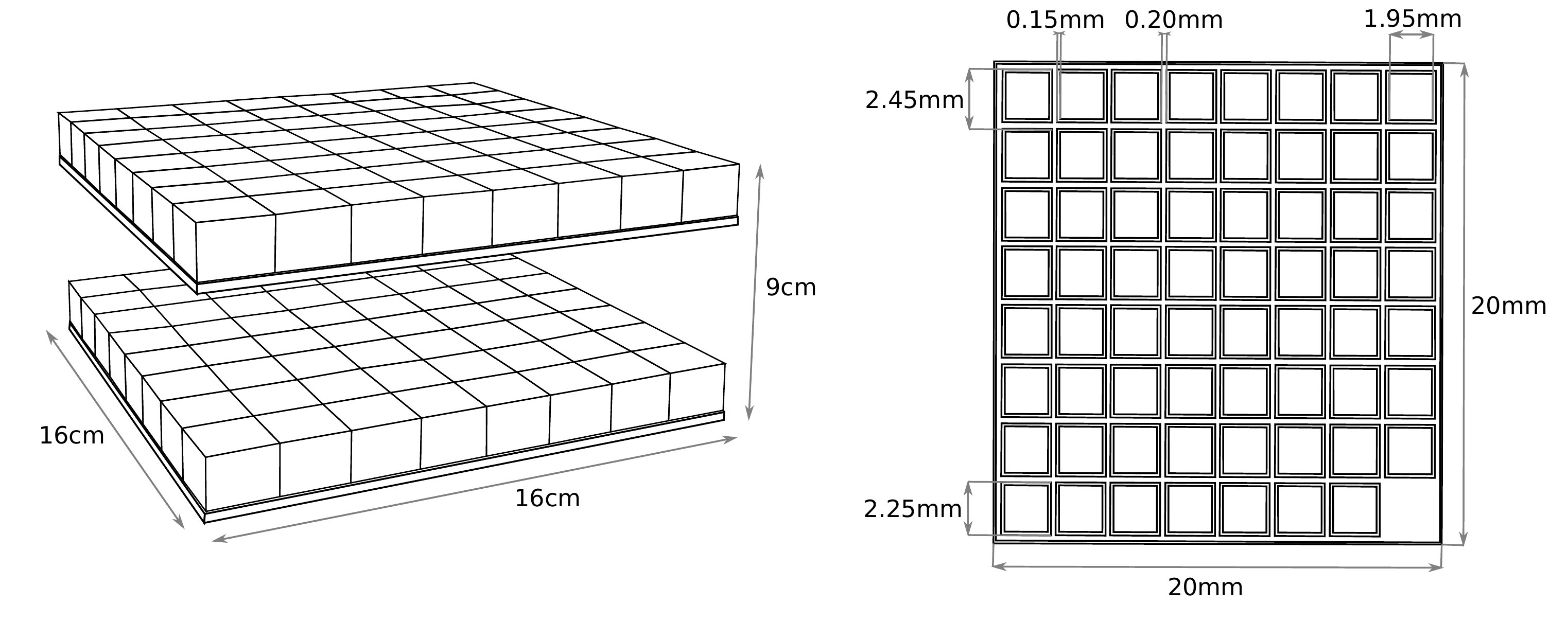}
\end{tabular}
\end{center}
\caption[CubeSat_Geo]{Schematic model of MeVCube (left) and anode pattern of CdZnTe detectors (right).}
\label{fig:CubeSat_Geo}
\end{figure}

\noindent \emph{MeVCube} is a 6U CubeSat concept currently under investigation at \emph{DESY}, with a scientific payload that can fit in an estimated volume of $20 \times 20 \times 10\;\mathrm{cm}^3$ (see Fig.~\ref{fig:CubeSat_Geo}). The ``core'' of the instrument consists of $128$ Cadmium-Zinc-Telluride (CdZnTe) detectors, arranged on two layers. Soft gamma-rays in the MeV energy range predominately interact through Compton scattering in the first layer (the scattering zone) and the scattered photon is absorbed in the second layer (the absorption zone). In this configuration the direction of the incoming photons can be confined to thin circles on the sky \cite{Ballmoos_ComptonImaging}. More specifically, the arrival direction of gamma-rays is determined by the measurements of the scattered photon direction and the Compton angle, computed trough the electron and photon energies. Therefore detectors employed on a Compton camera need both good energy and spatial resolution. Cadmium-Zinc-Telluride semiconductor detectors are very attracting for this kind of application, due to the high atomic number, density, wide band gap and low leakage current, that ensure high detection efficiency and good spectral and imaging performance at room temperature \cite{Schlesinger_CZTreview_2001}. In our design, each CdZnTe detector has a volume of $2.0\times 2.0 \times 1.5\;\mathrm{cm}^3$ and an $8\times8$ pixel structure ($2.45\;\mathrm{mm}$ pitch). The triggered pixel signal gives information on the energy deposited in the detector and the x-y location, while the interaction depth (z location) is reconstructed from the ratio between the cathode and pixel signals.\\
CdZnTe detectors are read-out by VATA450.3 ASICs, developed by \textit{Ideas}. VATA450.3 shows promising performances in terms of dynamic range, noise and linearity, while its low-power design allows to operate the detectors under the power constrains present in small satellites.\\
The Compton camera will be enclosed by an \textit{anti-coincidence detector} (ACD), composed of $5$ slabs of light plastic scintillator, read by silicon photo-multipliers (not shown in Fig.~\ref{fig:CubeSat_Geo}). Performance and reliability of such detectors have already been deeply investigated by previous missions, like \textit{Fermi}-LAT. The main \emph{MeVCube} specifications and requirements are summarized in Tab.~\ref{tab:MeVCUbeSpecifications}.\\

\begin{table}[ht]
\begin{center}       
\begin{tabular}{|l|l|}
\hline
\rule[-1ex]{0pt}{3.5ex} \textbf{Parameter} & \textbf{Value} \\
\hline
\rule[-1ex]{0pt}{3.5ex} CubeSat model & 4U scientific payload, \\
 & 6U complete satellite \\
\hline
\rule[-1ex]{0pt}{3.5ex} Orbit & Low Earth Orbit (LEO), $\sim 550\;\mathrm{km}$ \\
\hline
\rule[-1ex]{0pt}{3.5ex} Number of CdZnTe detectors & $128$ \\
\hline
\rule[-1ex]{0pt}{3.5ex} CdZnTe detector size & $2.0\;\mathrm{cm} \times 2.0\;\mathrm{cm} \times 1.5\;\mathrm{cm}$ \\
\hline
\rule[-1ex]{0pt}{3.5ex} Pixel pitch & $2.45\;\mathrm{mm}$ \\
\hline
\rule[-1ex]{0pt}{3.5ex} Depth resolution & $\lesssim 2.0\;\mathrm{mm}$ (FWHM) \\
\hline
\rule[-1ex]{0pt}{3.5ex} Energy resolution & $\lesssim 3.0\,\%$ at $662\;\mathrm{keV}$ (FWHM) \\
\hline
\rule[-1ex]{0pt}{3.5ex} Read-out electronics & VATA450.3 \\
\hline
\rule[-1ex]{0pt}{3.5ex} Total power consumption & $<5\;\mathrm{W}$ \\
\hline
\end{tabular}
\end{center}
\caption{MeVCube specifications.} 
\label{tab:MeVCUbeSpecifications}
\end{table}

\begin{figure}[ht]
\begin{center}
\begin{tabular}{c} 
\includegraphics[width=0.70\columnwidth]{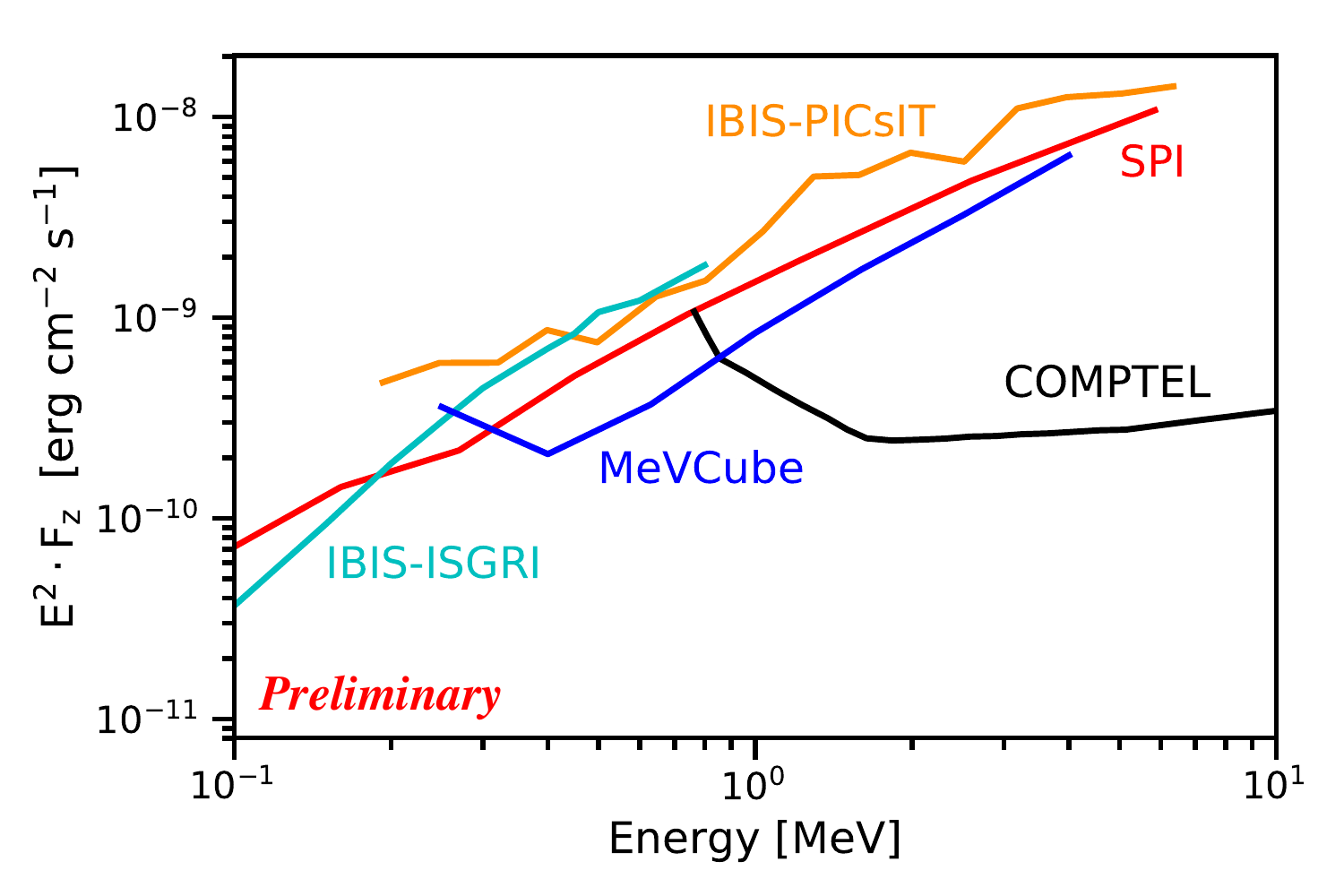}
\end{tabular}
\end{center}
\caption[CubeSat_Sensitivity]{Preliminary \emph{MeVCube} sensitivity estimation for a $10^5\;\mathrm{s}$ observation of sources at high latitude, compared to the ones achieved by \emph{COMPTEL} \cite{Comptel_Sens} and by the IBIS and SPI instrument on-board the \emph{INTEGRAL} observatory \cite{SPI_Sens, IBIS_Sens}.}
\label{fig:CubeSat_Sensitivity}
\end{figure}

\noindent \emph{MeVCube} response is evaluated through the \textit{Geant4} based simulation toolkit \textit{MegaLib} \cite{Zoglauer_MEGAlib}. The continuum sensitivity, which quantifies the telescope's ability to detect faint sources in presence of background, is calculated based on instrument performance like background rate, effective area, observation time and angular resolution, through the following formula:
\begin{equation}
F_{z} = \frac{z^2 + z \sqrt{z^2 + 4 N_{\mathrm{bkg}}}}{2 T_{\mathrm{obs}} A_{\mathrm{eff}}} \, .
\label{eq:sensitivity} 
\end{equation}
Here $z$ is the statistical significance in unit of sigmas (here $3\sigma$), $T_{\mathrm{obs}}$ is the observation time, $A_{\mathrm{eff}}$ the effective area and $N_{\mathrm{bkg}}$ is the number of background events that lie inside the angular resolution element defined by the telescope. A review of the background sources for a Compton telescope in a low-Earth orbit is reported in Ref.~\citenum{Cumani_Background_2019}. The main background contributions come from the extra-galactic background, and the Earth's albedo photons generated from the interaction of cosmic rays with the upper layers of the atmosphere. On the other hand, charged-particle background can be effectively vetoed by the ACD. \emph{MeVCube} sensitivity computed for gamma-ray sources at high latitude and for an observation time of $10^5\;\mathrm{s}$, assuming the specifications of Tab.~\ref{tab:MeVCUbeSpecifications}, is plotted in Fig.~\ref{fig:CubeSat_Sensitivity}: the telescope can cover the energy range between $200\;\mathrm{keV}$ and $4\;\mathrm{MeV}$ with a sensitivity comparable to the one reached by \emph{COMPTEL} or \emph{INTEGRAL} satellites.

\section{Experimental results of a custom design CdZnTe detector}
\label{sec:results}
\noindent Performance of a $2.0\;\mathrm{cm} \times 2.0 \;\mathrm{cm} \times 1.5 \;\mathrm{cm}$ pixelated CdZnTe detector, biased at $- 2500\;\mathrm{V}$, was measured using different radioactive sources. Pixels signals are read-out by VATA450.3 ASIC on an evaluation testboard directly provided by \textit{Ideas}. A description of the operating principle of the ASIC and measurements of its performance can be found in Ref.~\citenum{Lucchetta_SPIE}. A preliminary read-out system for the cathode, based on a discrete AMPTEK A250 charge sensitive pre-amplifier\footnote{https://www.amptek.com/internal-products/a250-charge-sensitive-preamplifier} and a DRS4 ASIC\footnote{https://www.psi.ch/en/drs/evaluation-board}, has been developed at DESY \footnote{on a later stage of the project the entire read-out will be based on VATA450.3.}.\\

\begin{figure}[ht]
\centering
\begin{subfigure}[b]{.45\textwidth}
\hspace{0.1cm}
\includegraphics[width=0.99\columnwidth]{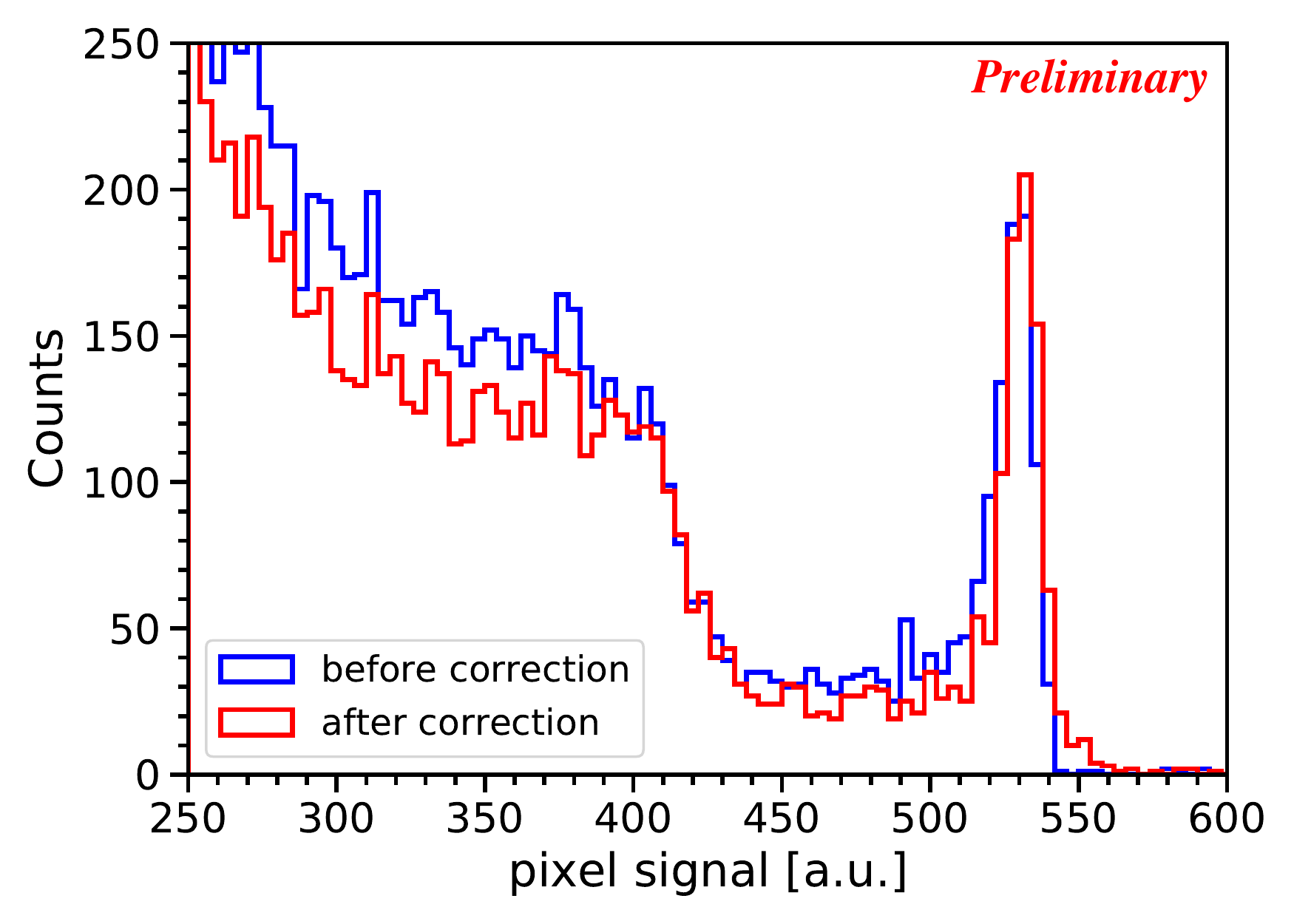}
\vspace{2ex}
\includegraphics[width=0.99\columnwidth]{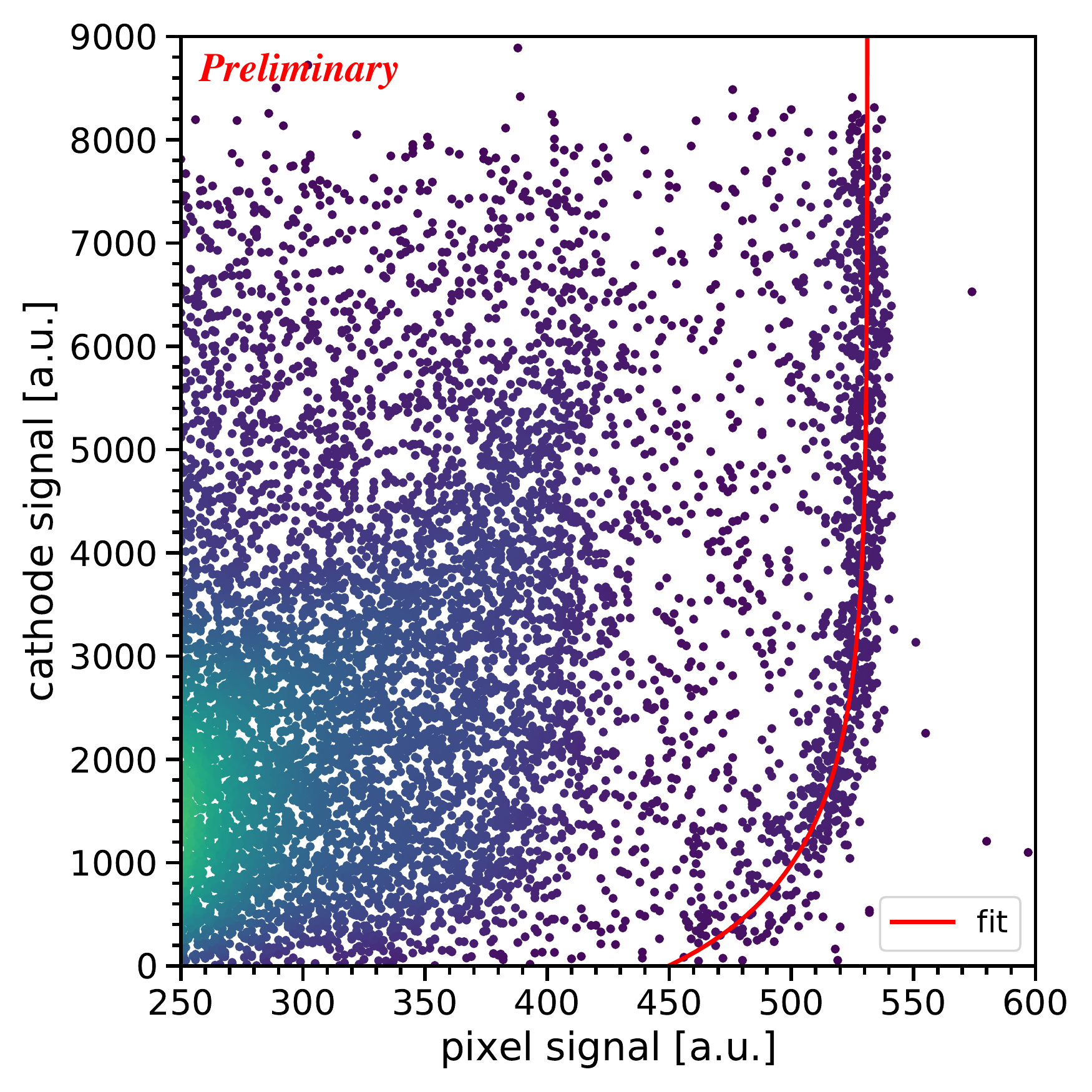}
\end{subfigure}
\qquad
\begin{subfigure}[b]{.45\textwidth}
\hspace{0.1cm}
\includegraphics[width=0.99\columnwidth]{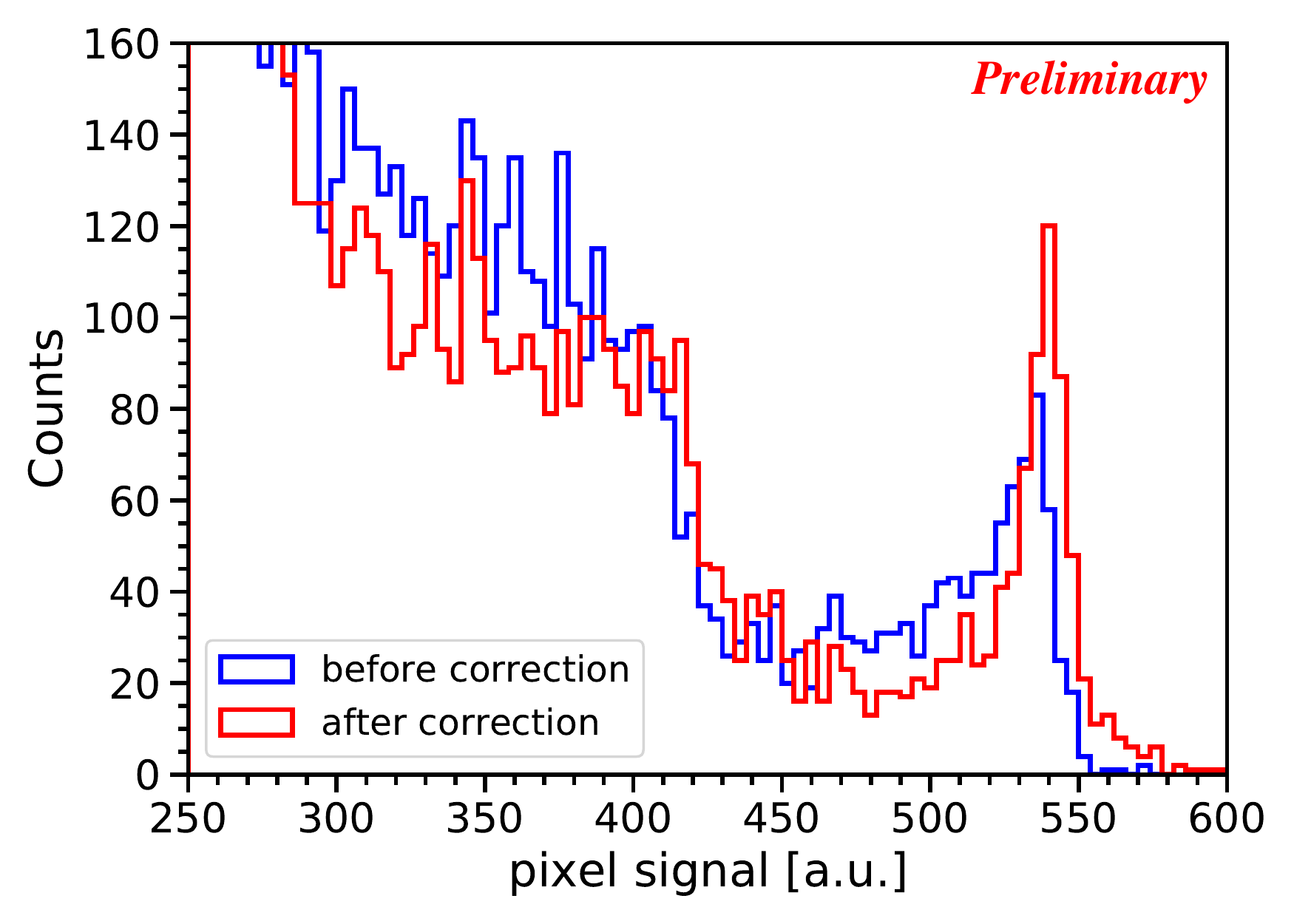}
\vspace{2ex}
\includegraphics[width=0.99\columnwidth]{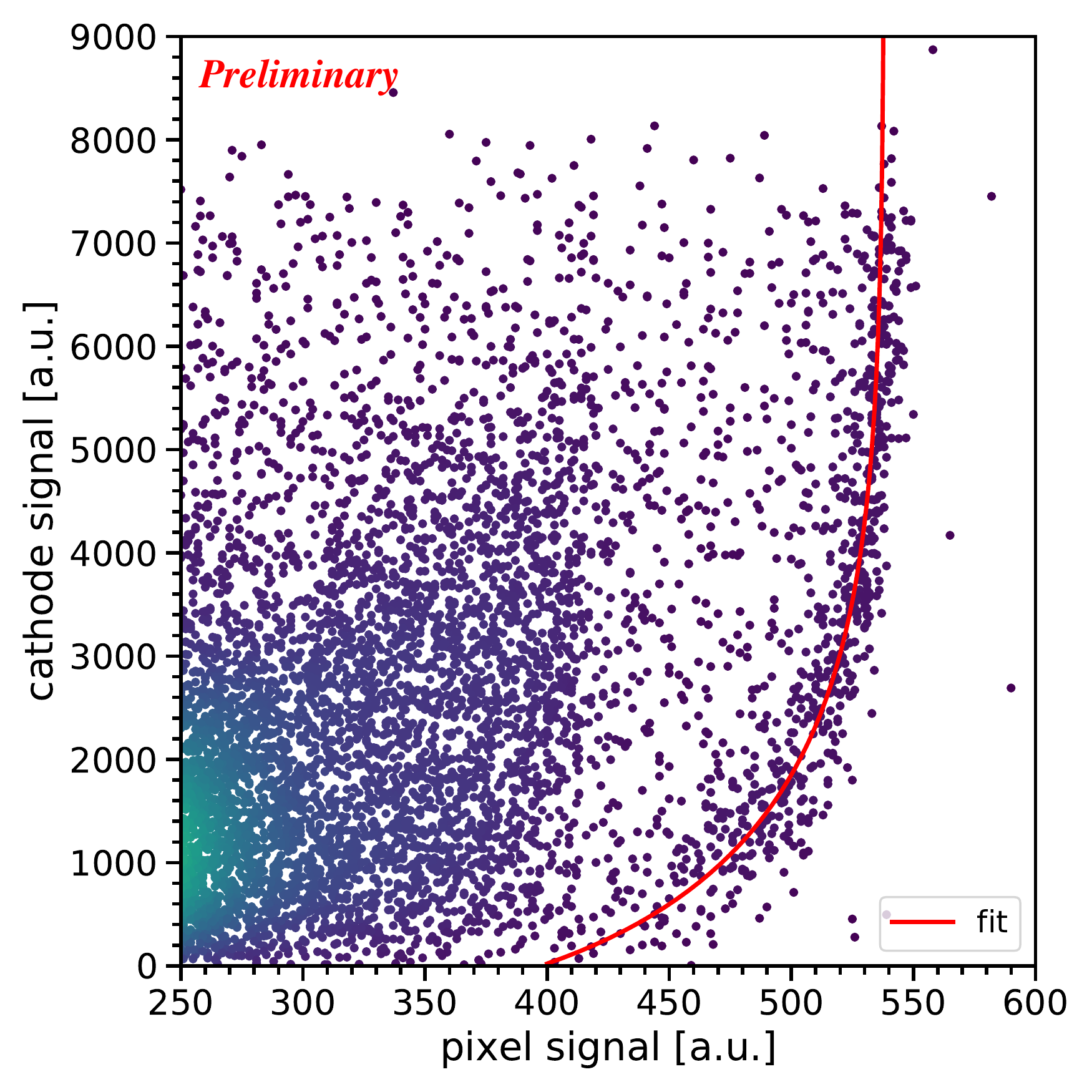}
\end{subfigure}
\caption[Doi_technique]{Energy spectra from a $^{137}Cs$ source for a couple of selected pixels, and illustration of the \textit{depth-of-interaction} technique. Signals from the cathode, taken in coincidence with those of the pixels, provide a correction for incomplete charge collection and trapping.}
\label{fig:Doi_technique}
\end{figure}

\noindent In thick CdZnTe detectors the measured pixel spectra might exhibit pronounced left tails due to incomplete charge collection and trapping effect (see blue histograms in Fig.~\ref{fig:Doi_technique}). However a distinct correlation profile exists between signals from the cathode and those of the segmented anode electrodes, as illustrated in the scatter plot in Fig.~\ref{fig:Doi_technique} (below). The linearization of these profiles, known in literature as \textit{depth-of-interaction} correction \cite{Shor_CZT}, provides an overall improvement of the spectral performance of the detector, resulting in sharper energy lines for the photopeaks (red histograms in Fig.~\ref{fig:Doi_technique}).

\begin{figure}[ht]
\centering
\begin{subfigure}[b]{.57\textwidth}
\includegraphics[width=0.99\columnwidth]{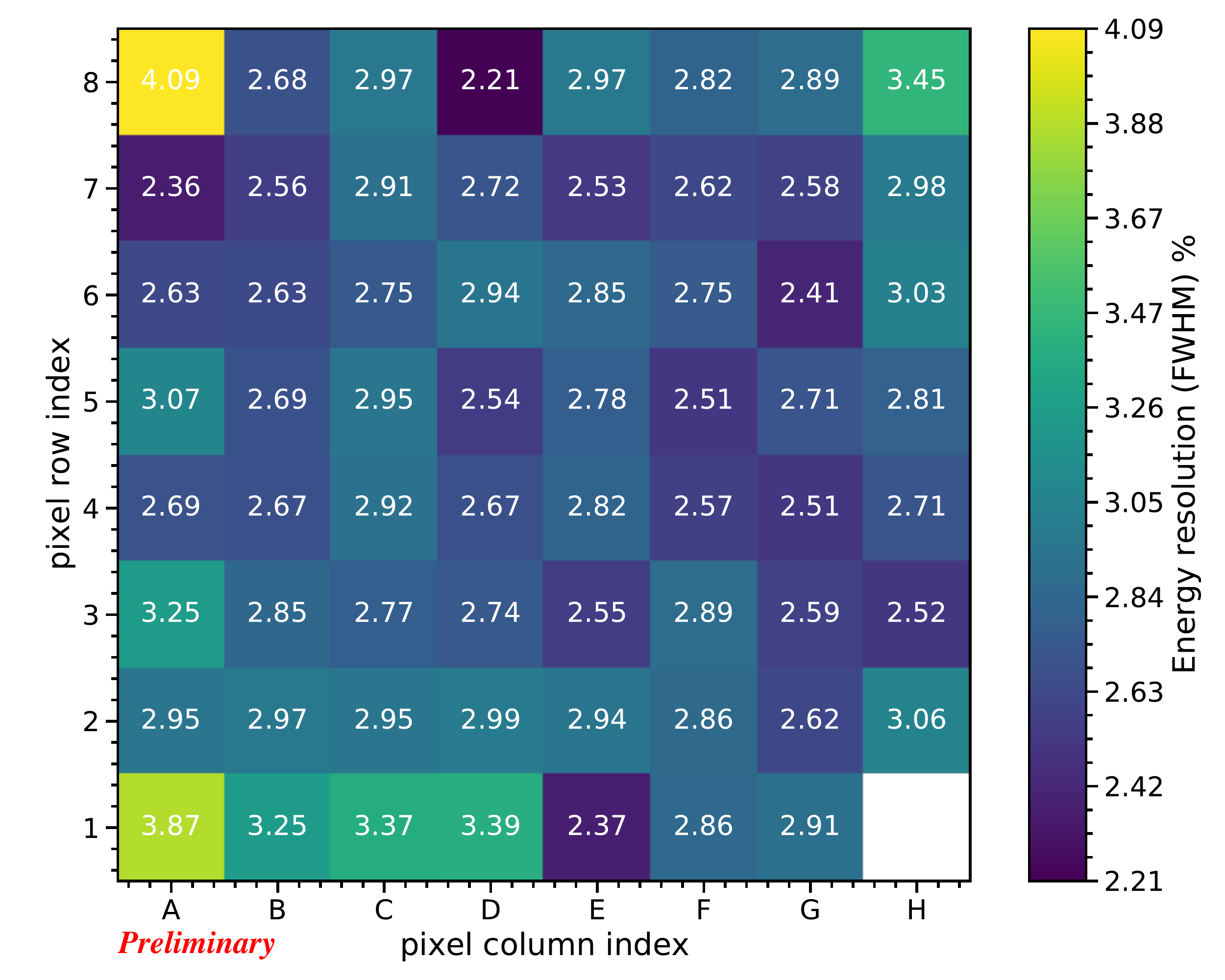}
\end{subfigure}
\qquad
\begin{subfigure}[b]{.37\textwidth}
\includegraphics[width=0.92\columnwidth]{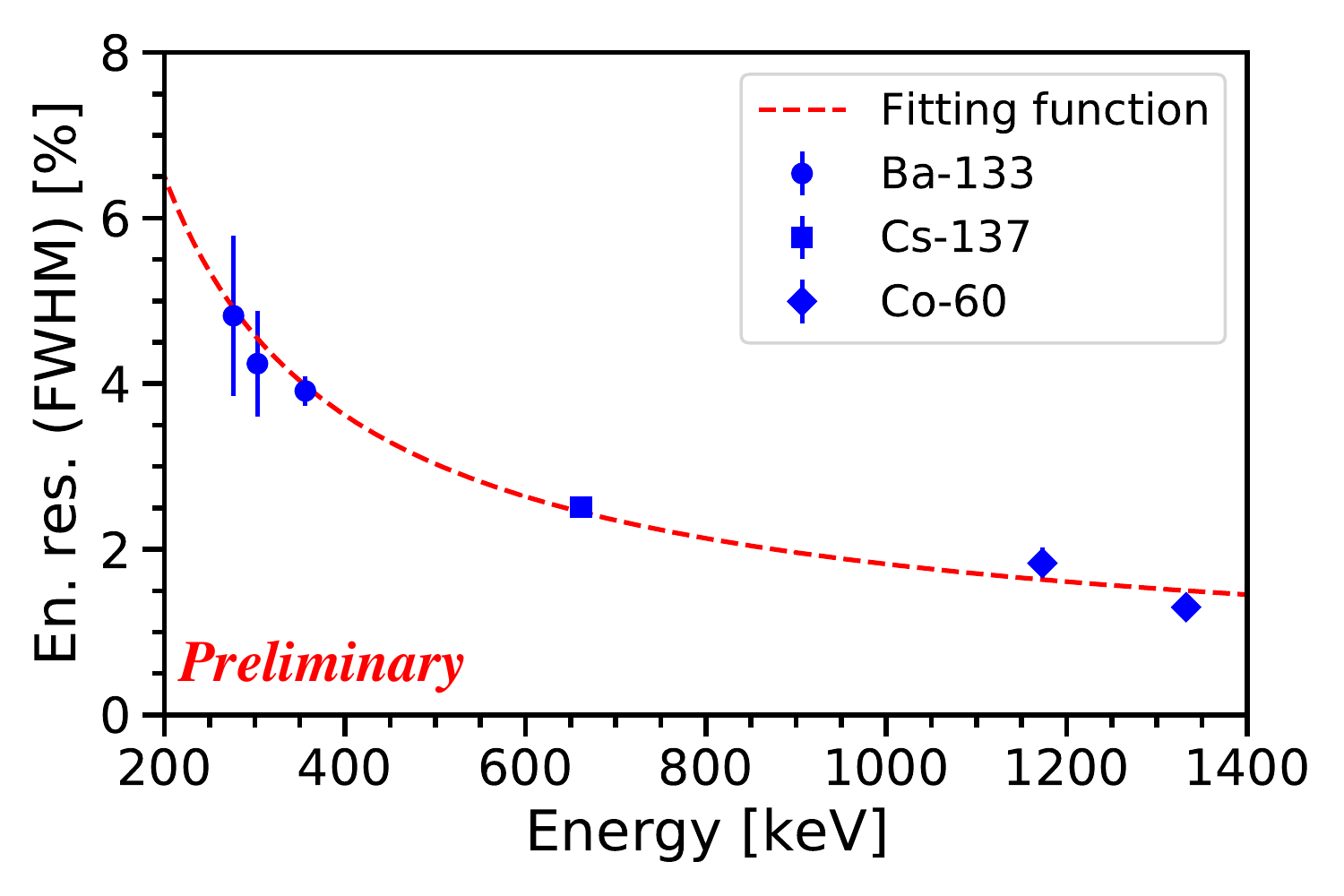}
\includegraphics[width=0.92\columnwidth]{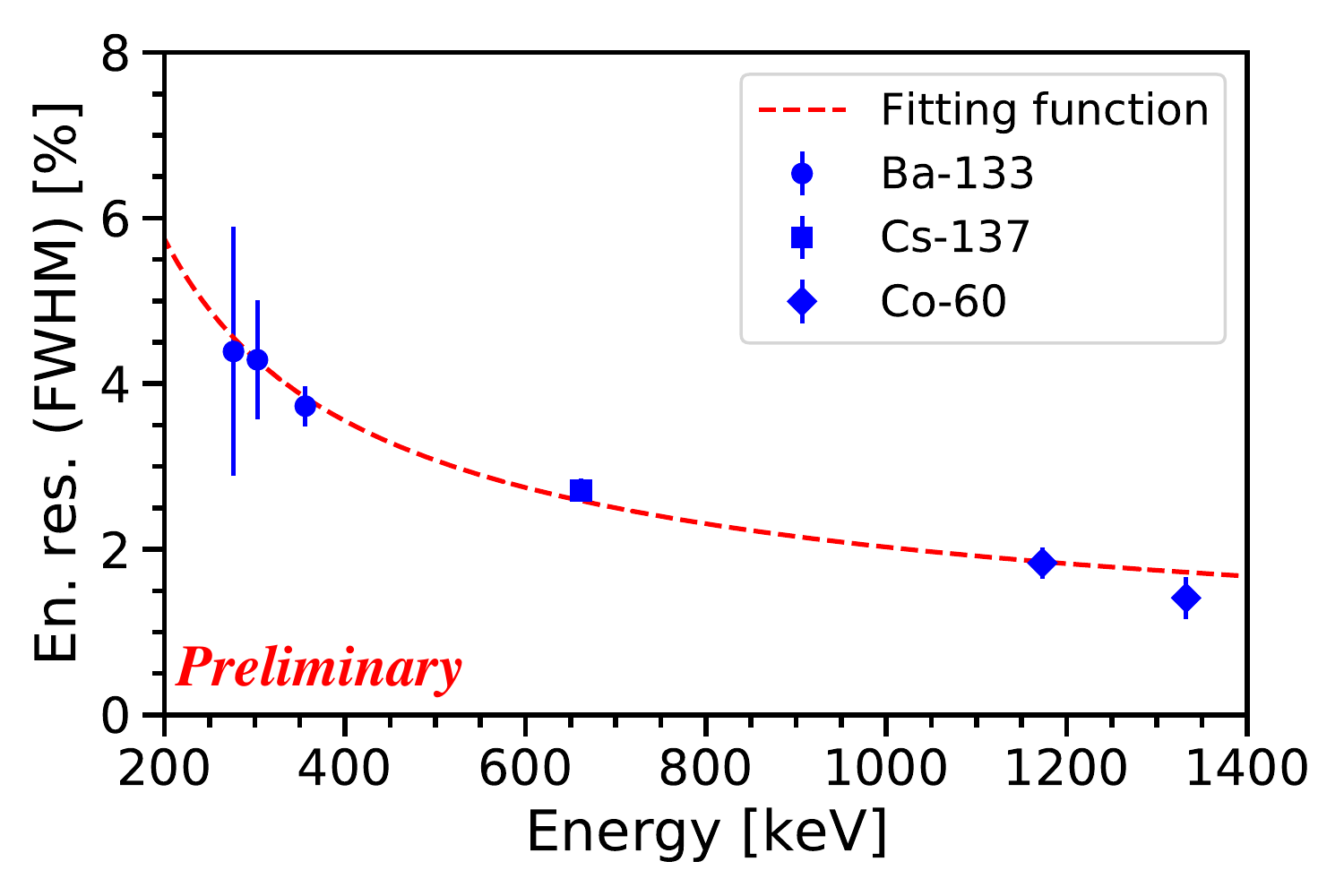}
\end{subfigure}
\caption[EnergyResolution]{Energy resolution (expressed in FWHM) measured for the $662\;\mathrm{keV}$ line of $^{137}Cs$ for all pixels (left), and with different radioactive sources for a couple of selected pixels (right).}
\label{fig:EnergyResolution}
\end{figure}

\noindent Energy resolution measurements of the prototype detector are shown in Fig.~\ref{fig:EnergyResolution}. The left plot depicts the energy resolution for all pixels, expressed in full width at half maximum (FWHM), for a $^{137}Cs$ radioactive source. The spectral performance is very uniform throughout the detector, with majority of pixels showing an energy resolution $\lesssim 3.0\%$ at $662\;\mathrm{keV}$. A slightly worse energy resolution is measured for few edge pixels, reasonably due to imbalances of electric field in the boundaries. The spectral performance measured for different radioactive sources is reported in the right plots of Fig.~\ref{fig:EnergyResolution}, for a couple of selected pixels: the measured energy resolution is $\sim 6\%$ around $200\;\mathrm{keV}$ and lowers to $\lesssim 2\%$ at energies $>1\;\mathrm{MeV}$.\\
\newline

\begin{figure}[ht]
\centering
\begin{subfigure}[b]{.47\textwidth}
\includegraphics[width=0.99\columnwidth]{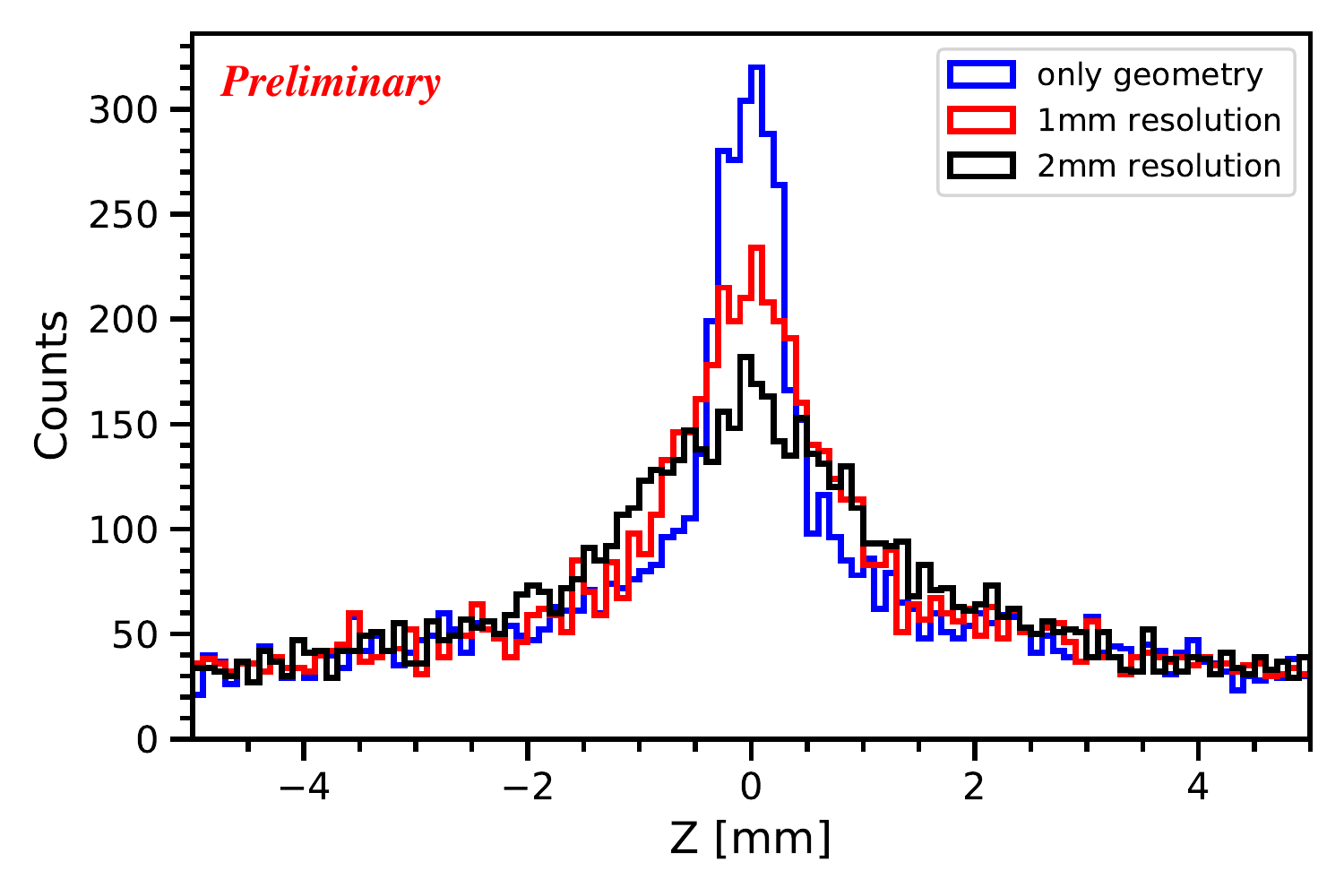}
\end{subfigure}
\qquad
\begin{subfigure}[b]{.47\textwidth}
\includegraphics[width=0.99\columnwidth]{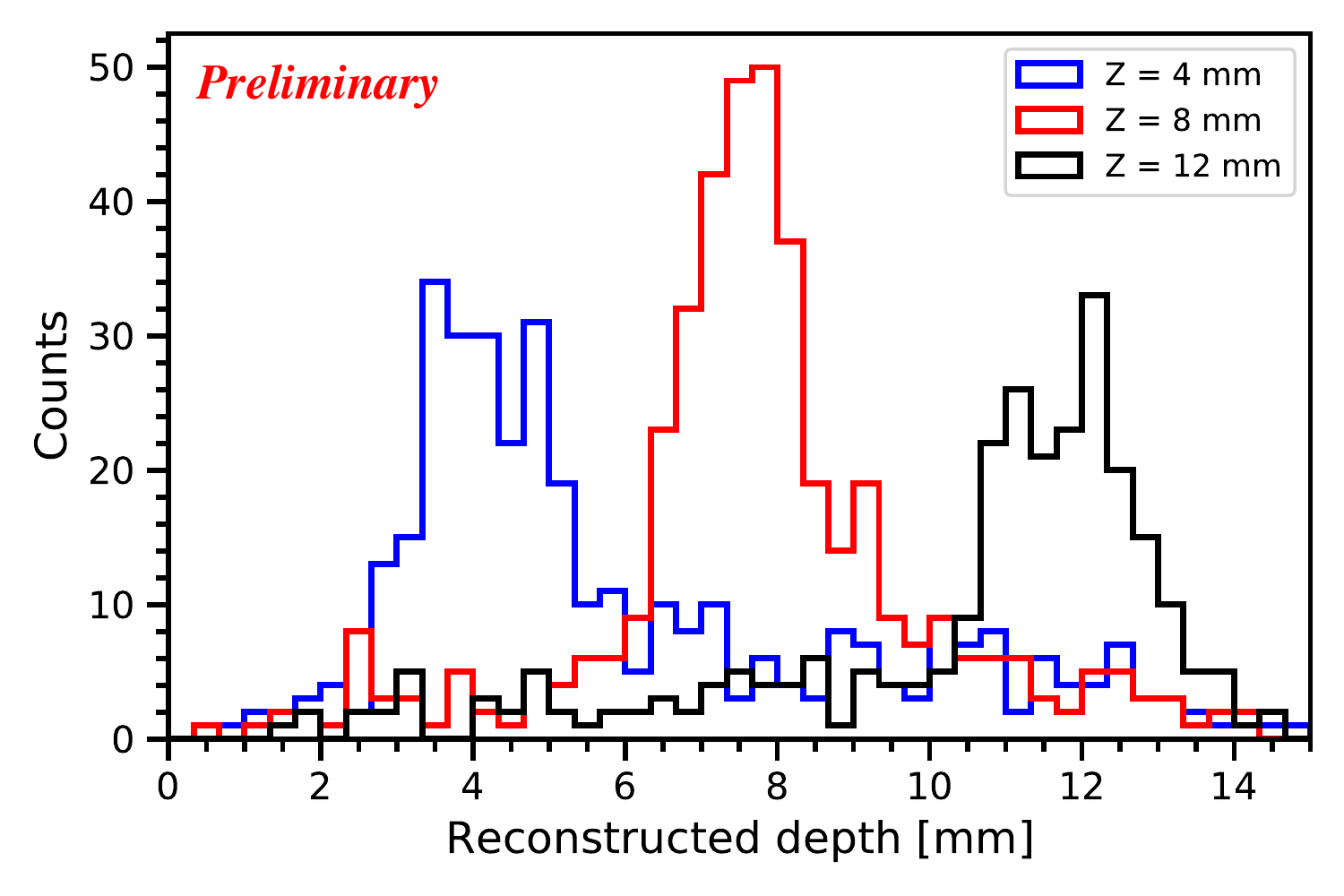}
\end{subfigure}
\caption[DepthResolution]{Interaction depth obtained from a \textit{Geant4} simulation (left) and measured one for three different scanning positions (right).}
\label{fig:DepthResolution}
\end{figure}

\noindent The spatial resolution of the detector was investigated with a $^{137}Cs$ source and a copper collimator. The collimator has a length of $10\;\mathrm{cm}$ and a drill hole of $0.5\;\mathrm{mm}$ in diameter. The collimator stands to a distance of $2.8\;\mathrm{cm}$ from the detector. The left plot in Fig.~\ref{fig:DepthResolution} reports the depth resolution obtain from a \textit{Geant4} \cite{Geant4} simulation considering a CdZnTe detector with infinitely precise energy and spatial resolution (thus evaluating only the ``geometrical'' effects, due to the finite size of the collimator beam), and when smearing the data with $1\;\mathrm{mm}$ and $2\;\mathrm{mm}$ resolution (FWHM). The gaussian profiles exhibit pronounced tails due to incomplete photon absorption and effect of the collimator penumbra. The measured depth resolution for three different scanning positions is shown in the right of Fig.~\ref{fig:DepthResolution}. The interaction depth is estimated selecting events corresponding to the Cesium photo-peak and computing the ratio between the cathode and the anode signals. The widths of the distributions are in the range between $1.7-1.9\;\mathrm{mm}$ in FWHM, implying a spatial resolution of $\sim 1.5-1.7\;\mathrm{mm}$, when subtracting the geometrical component evaluated trough the simulations.

\section{Conclusions}
\noindent In this manuscript, the viability and the performance of a Compton telescope, based on the CubeSat standard, was evaluated. \emph{MeVCube} can operate in the energy range between hundreds keV up to few MeVs, with a sensitivity comparable to the one achieved by \emph{COMPTEL} and \emph{INTEGRAL}. Therefore, even a small Compton camera, flying on a CubeSat, will not only be a technology demonstrator, but can have its own scientific impact. Experimental measurements on a custom design CdZnTe detector have shown promising results in terms of energy and spatial resolution, validating the requirements imposed in the simulations. Specifically, the energy resolution ranges from $\sim 6\%$ around $200\;\mathrm{keV}$ to $\lesssim 2\%$ above $1\;\mathrm{MeV}$, while a depth resolution around $1.5-1.7\;\mathrm{mm}$ was obtained.

%
%
%

\end{document}